\documentclass[3p,times,twocolumn]{elsarticle}

\usepackage{ecrc}

\volume{00}

\firstpage{1}

\journalname{Nuclear Physics B Proceedings Supplement}

\runauth{}

\jid{nuphbp}

\jnltitlelogo{Nuclear Physics B Proceedings Supplement}

\usepackage{amssymb}
\usepackage[figuresright]{rotating}

\input babarsym

\begin{document}

\begin{frontmatter}



\dochead{}

\title{Update of the Hadronic Vacuum Polarisation Contribution to the muon g-2}


\author{Michel Davier}

\address{LAL, CNRS/IN2P3 et Universit\'e Paris-Sud 11, 91898 Orsay, France}

\begin{abstract}
Precise data on $e^+e^-\to {\rm hadrons}$ have recently become available and are 
used to compute the lowest-order hadronic vacuum polarisation contribution to
the muon magnetic anomaly through dispersion relations. This is the case for the
dominant $\pi^+\pi^-$ channel, but the most significant progress comes from the near
completion of the \babar\ program of measuring exclusive processes below 2 GeV with 
the initial-state radiation method which allows an efficient coverage of a large
range of energies.. In this paper we briefly review the data treatment, the
achieved improvements, and the result obtained for the full Standard Model prediction
of the muon magnetic anomaly. The value obtained,
$a_\mu^{\rm had~LO}=(692.6 \pm 3.3)\times 10^{-10}$ is 20\% more precise than our 
last estimate in 2010. It deviates from the direct experimental determination
by $(27.4 \pm 7.6)\times 10^{-10}$ (3.6$\sigma$). Perpectives for further 
improvement are discussed. 
\end{abstract}


\end{frontmatter}

% \linenumbers

\section{ Hadronic vacuum polarisation to muon $g-2$ and $e^+e^-$ data}
\label{HVP-intro}
The dominant part of the uncertainty in the Standard Model prediction for the 
muon magnetic anomaly $a_\mu=(g-2)/2$, where $g$ is the gyromagnetic ratio 
equal to 2 at the lowest QED order, comes from the contribution of the lowest-order
(LO) hadronic vacuum polarisation (HVP). The latter is
computed through a dispersion relation using the measured
cross sections for $e^+e^-\to {\rm hadrons}$, as the relevant energy
scale is too low for applying perturbative QCD. The HVP component is given by:
\begin{equation}
\label{integral}
   a_\mu^{\rm had~LO}=\frac {1}{4\pi^3}\!\!
    \int_{m_\pi^2}^\infty\!\!ds\,K(s)\,\sigma^{0}_{\rm hadrons}(s)~,
\end{equation}
where $K(s)$ is a QED kernel and $\sigma^{0}_{\rm hadrons}(s)$ the bare
cross section including final state radiation. Therefore progress on the 
HVP contribution is completely controled by the availability of precise and 
reliable data on the hadronic annihilation cross sections.

Since our last update in 2010~\cite{dhmz2010} (see also Ref.~\cite{hlmnt2011}) 
new experimental data became available. In particular the \babar\ collaboration 
has essentially completed a program of precise 
measurements of exclusive cross sections for all the dominant channels of 
$\epem\to {\rm hadrons}$ from threshold to an energy of 3-5 GeV using 
the initial-state radiation (ISR) method. Also results are being produced at the
VEPP-2000 facility in the 1-2 GeV range. In this paper we present our improved 
prediction using these new input data. 

\section{Data treatment}
\label{treatment}

Our procedure for computing the dispesion relation has evolved with several new 
ideas to improve precision and reliability. At a time when the quality of 
$e^+e^-$ data was limited we proposed in 1997 to use instead data from hadronic 
$\tau$ decays assuming CVC and taking into account isospin-breaking 
effects~\cite{adh}, taking advantage of
the pure $\tau$ decay sample in the ALEPH experiment. 
Furthermore the relative normalization with respect to the $\tau$ leptonic 
decay was known very precisely from the measurement of the branching 
ratios. Through detailed QCD studies the $\tau$ hadronic spectral functions were
shown to be well described by quark-hadron duality~\cite{svz} so that one could 
propose and justify in 1998 to use perturbative QCD at energies as low as the $\tau$ 
mass\cite{dh98-1,dh98-2}. With the availability of VEPP-2M more precise data a
substantial update was published in 2003~\cite{dehz2003-1,dehz2003-2}. Unmeasured 
channels were estimated or bounded using isospin constraints. In 2010 a more
detailed study of isospin breaking when using $\tau$ data was 
performed~\cite{DHLMMTWYZ}, yielding 
better agreement with $e^+e^-$ data. Reliability was increased with improved
statistical and systematic tools for the treatment and combination of the data
from different experiments through the package HVPTools, and the \babar\ 
$\pi^+\pi^-$ was included~\cite{dhmyz}. Finally the last global update using all
the available $e^+e^-$ data in 2010 produced the value~\cite{dhmz2010}~
\footnote{When not explicitely noted, all $a_\mu^{\rm had}$ values quoted in this
paper are in units of $10^{-10}$.}

\begin{equation}
\label{res-2010}
 a_\mu^{\rm had~LO}=(692.3 \pm 4.2)\times 10^{-10}~~~~~~~(2011)~.
\end{equation}

Experimental exclusive cross sections are integrated using Eq.~(\ref{integral}) up
to 1.8 GeV.
In the present work 37 exclusive channels are included, as compared to only 22 
in 2010. Thanks to the larger completeness of the data sample, only very few 
channels are now estimated with isospin constraints. In the energy range 1.8-3.7
GeV and above 5 GeV 4-loop perturbative QCD is used~\cite{baikov}. The contribution
from the open charm
region 3.7-5 GeV is again computed with experimental data. The narrow resonances
$J/\psi$ and $\psi(2S)$ Breit-Wigner line shapes are integrated using their
currently best known parameters. 

The integration of data points belonging to different experiments with their own
data densities requires a careful treatment, especially concerning the 
correlated systematic uncertainties within the same experiment or between
different experiments using the same tools. Quadratic interpolation of the data
points is performed for each experiment and a local weighted average between the
interpolations is computed in 1-MeV bins. Full covariance matrices are 
constructed between experiments and channels. Errors are propagated using
pseudo-experiments (toys). When results from different experiments are 
locally inconsistent the
error is rescaled according to the $\chi^2$ value. At present, for the dominant 
$\pi^+\pi^-$ channel this is the major limiting factor for further improving the
precision. Except for very few energy regions in a couple of channels the 
largest weight in the combination is obtained for the \babar\ experiment.

\section{The dominant $\pi^+\pi^-$ channel}
\label{newdata}

\begin{figure}[htb]
  \centering
%  \vspace{-0.5cm}
  \includegraphics[width=7.5cm]{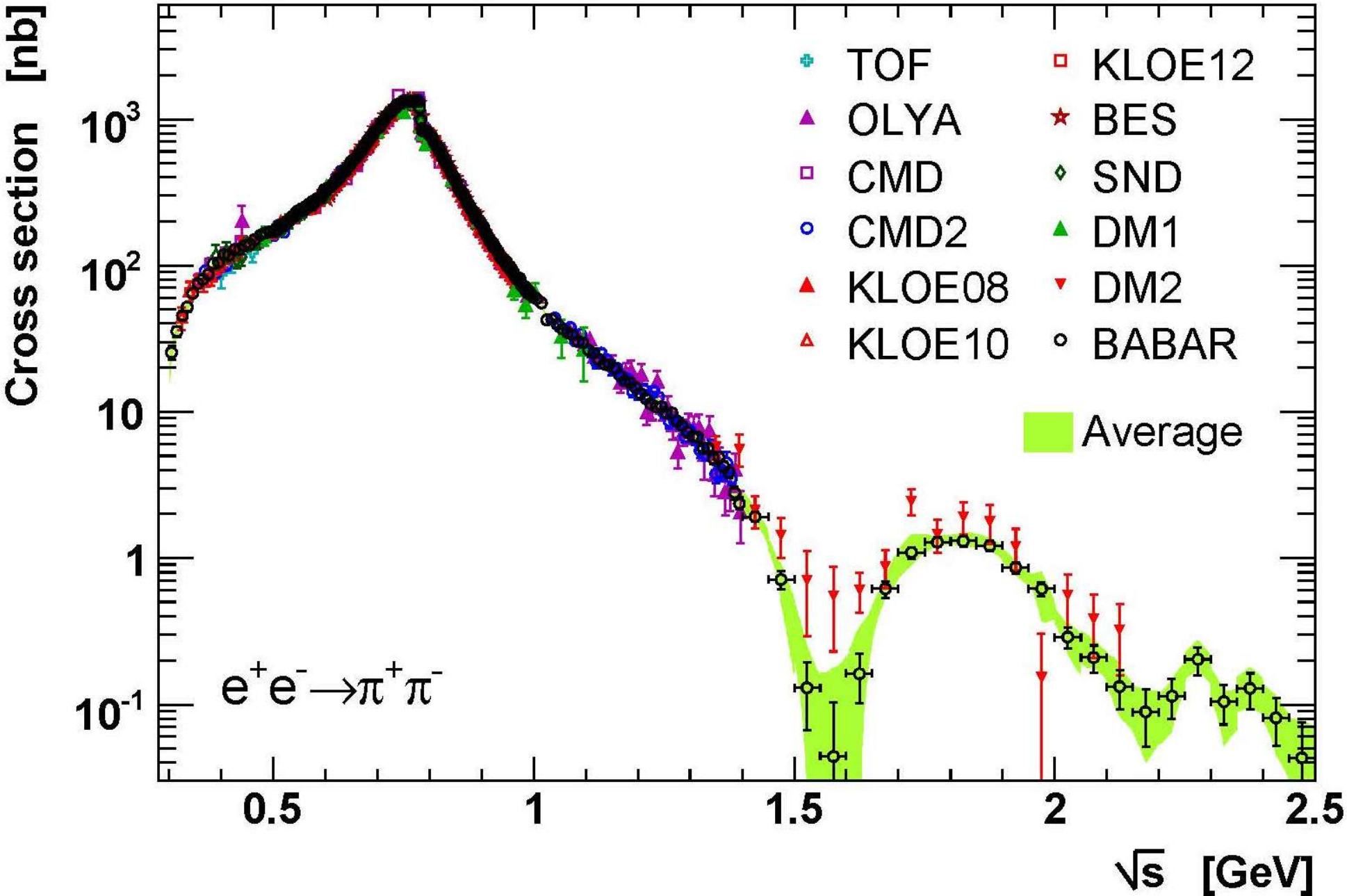}
  \vspace{-0.4cm}
  \caption{\small
The measured bare cross sections for $e^+e^-\to\pi^+\pi^-(\gamma)$  
with the combination band from threshold to 2.5 GeV. References are given in the text.}
  \label{pipiall}
\end{figure}

\begin{figure}[htb]
  \centering
%  \vspace{-0.5cm}
  \includegraphics[width=6cm]{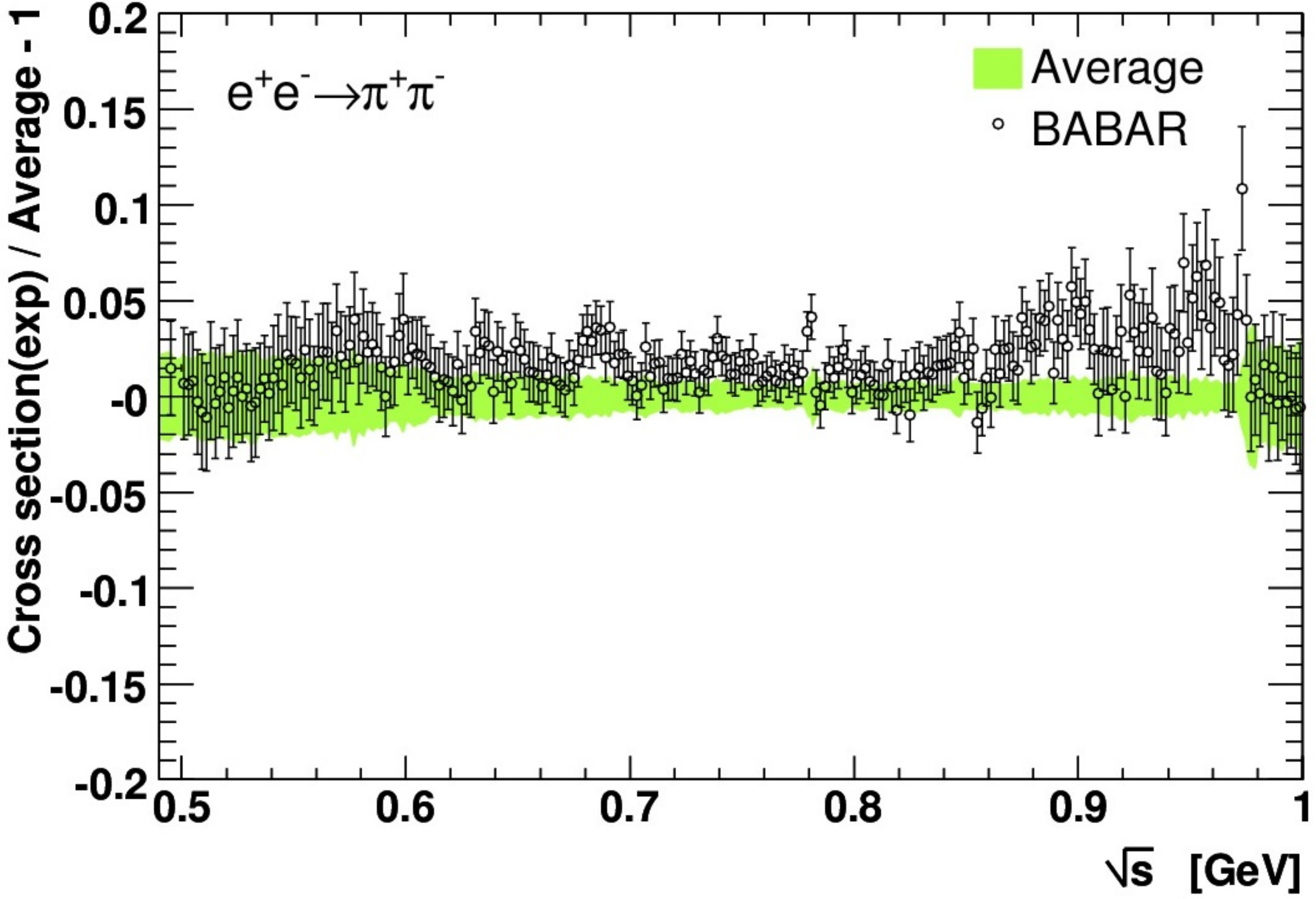}//
  \includegraphics[width=6cm]{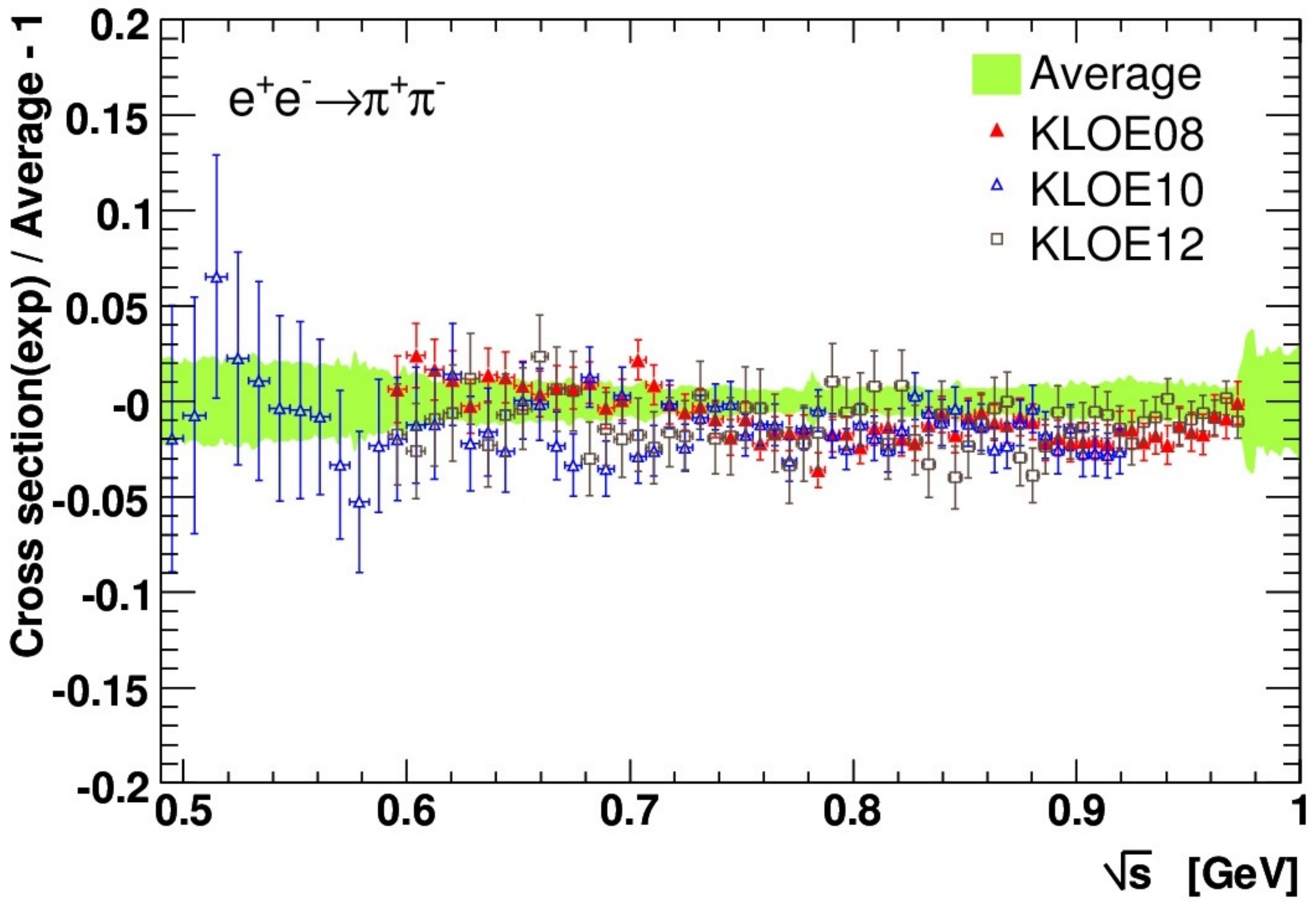}
  \vspace{-0.4cm}
  \caption{\small
The ratio of the measured cross sections for $e^+e^-\to\pi^+\pi^-(\gamma)$ to the
average for \babar\ (top) and the different KLOE measurements (bottom). } 
  \label{babar-kloe}
\end{figure}
The $\pi^+\pi^-$ channel dominates both the HVP contribution and its uncertainty.
Recent experiments are generally limited by systematic uncertainties 
$\delta_{\rm sy}$. 
The main older contributors (references are given in Ref.~\cite{dhmz2010} 
are CMD-2 ($\delta_{\rm sy}$=0.8\%),
SND (1.5\%), KLOE-2008 (0.8\%), KLOE-2010 (1.4\%), \babar\ (0.5\%). For this update
we included KLOE-2012~\cite{kloe2012} (0.8\%) and BESIII-2015~\cite{bes2015} 
(0.9\%). Only \babar\ covers the full
mass range with high precision. Overall the combination in Fig.\ref{pipiall} 
looks good. However a tension is observed between \babar\ and KLOE in the $\rho$
peak region (Fig.\ref{babar-kloe}), the other experiments falling in between and being
consistent with both. It is important to compare experiments at the cross section 
level where they should agree, rather than using the integrated values where local 
discrepancies could artificially cancel and not be included in the quoted systematic
uncertainty. Unhappily, the persisting discrepancy 
between \babar\ and KLOE leads to an enlargement of the combined uncertainty

In spite of this problem, the progress in estimating the $\pi^+\pi^-$ contribution
has been steady in the last decade. While the central value stayed within the
quoted uncertainties, the total uncertainty dropped from 5.9 in 2003 to 2.9 in
2010, and now 2.5. More precisely the updated value from threshold to 1.8 GeV is
$506.9 \pm 1.1_{\rm stat} \pm 2.2_{\rm uncor syst} \pm 0.7_{\rm cor syst}$, where the 
second (third) last uncertainty stands for systematic effects uncorrelated 
(correlated) with other channels, respectively. The correlation originates mainly 
from luminosity measurements and the VP correction.

Our last estimate~\cite{new-aleph-tau} using $\tau$ decay data from 
ALEPH, OPAL, CLEO, and Belle,
$516.2 \pm 2.9_{\rm exp} \pm 2.2_{\rm IB}$, where the second uncertainties arises 
from isospin-breaking (IB) corrections, is 2.2~$\sigma$ larger than the 
current $e^+e^-$-based value. 
The difference can be reduced by applying off-resonance
$\gamma-\rho$ mixing corrections~\cite{jeger-szafron} with additional 
uncertainties. Because of the impressive progress of $e^+e^-$ data, the $\tau$
input is now relatively less precise and less reliable due to the IB uncertainties.
While the $\tau$-$e^+e^-$ comparison is interesting in its own right, it is
safer now not to use it to evaluate precise HVP contributions.

\section{The four-pion channels}
\label{4pi}

\begin{figure}[htb]
  \centering
%  \vspace{-0.5cm}
  \includegraphics[width=7.5cm]{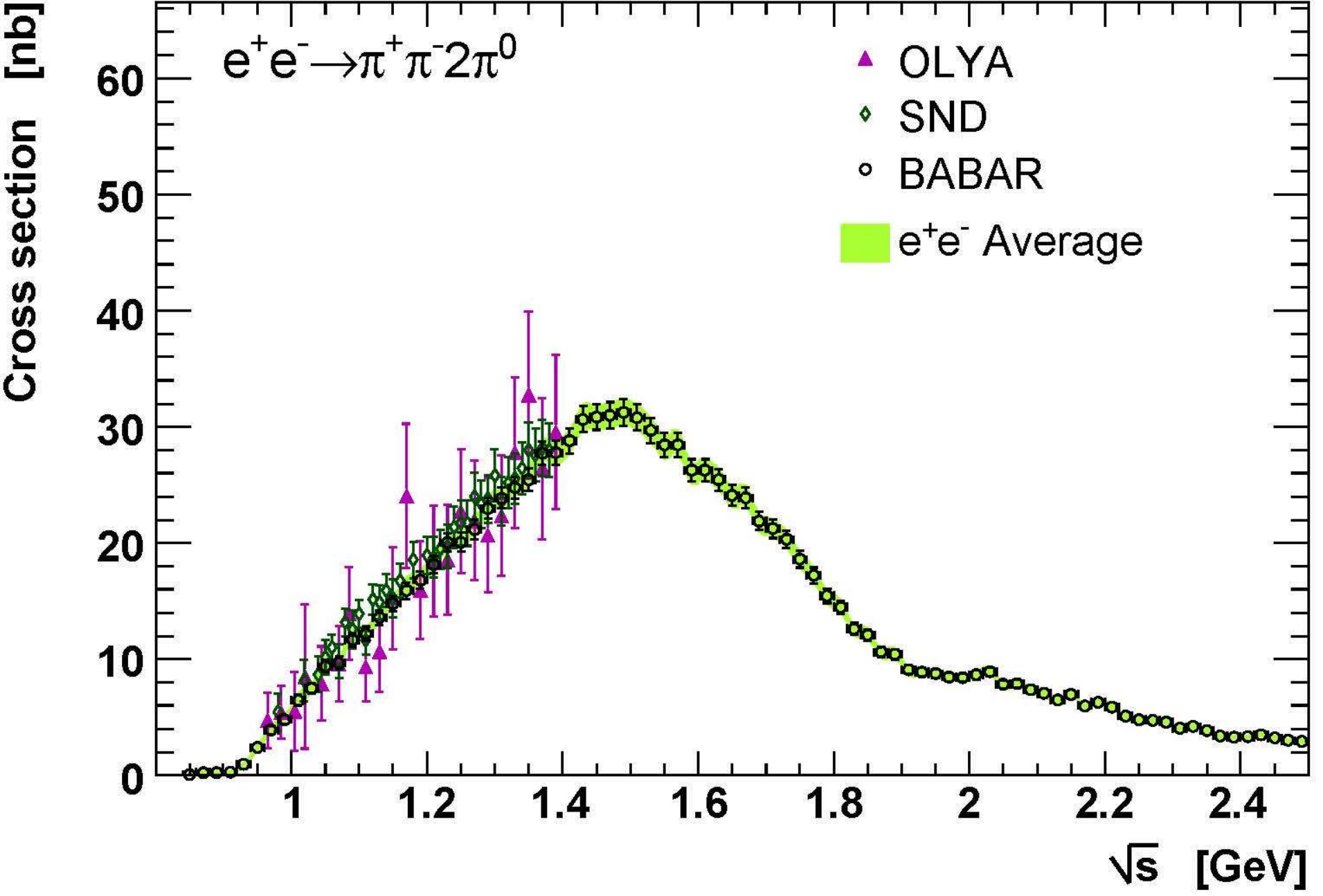}
  \vspace{-0.4cm}
  \caption{\small
The measured bare cross sections for $e^+e^-\to\pi^+\pi^-2\pi^0(\gamma)$  
with the combination band from threshold to 2.5 GeV.}
  \label{2pi2pi0}
\end{figure}

Preliminary results from \babar\ on $e^+e^-\to\pi^+\pi^-2\pi^0(\gamma)$ have been 
presented~\cite{2pi2pi0-babar}. As with other \babar measurements using the 
initial-state-radiation (ISR) method with the ISR photon detected at large angle,
the acceptance for the hadron system is large so that the final state structure
can be studied and taken into account in the Monte Carlo generator, hence reducing
significantly the systematic uncertainty on the acceptance. Data from some older 
experiments, both imprecise and inconsistent, are now discarded. As seen in
Fig.~\ref{2pi2pi0} the \babar\ results show a considerable improvement in 
precision.

The combined $\pi^+\pi^-2\pi^0$ HVP contribution from threshold to 1.8 GeV 
to $a_\mu$ comes out to be
$18.03 \pm 0.06_{\rm stat} \pm 0.49_{\rm uncor syst} \pm 0.26_{\rm cor syst}$ where
the total uncertainty 0.55 is much reduced compared to the 2011 value (1.24).
Recall the $\tau$ ALEPH estimate~\cite{new-aleph-tau} based on the 
$\nu_\tau \pi^{\pm} \pi^+ \pi^-$ and $\nu_\tau \pi^{\pm} 3\pi^0$ decay modes,
$21.02 \pm 1.16_{\rm exp} \pm 0.40_{\rm IB}$ which is 2.1~$\sigma$ larger,
albeit much less precise.

\begin{figure}[htb]
  \centering
%  \vspace{-0.5cm}
  \includegraphics[width=5cm]{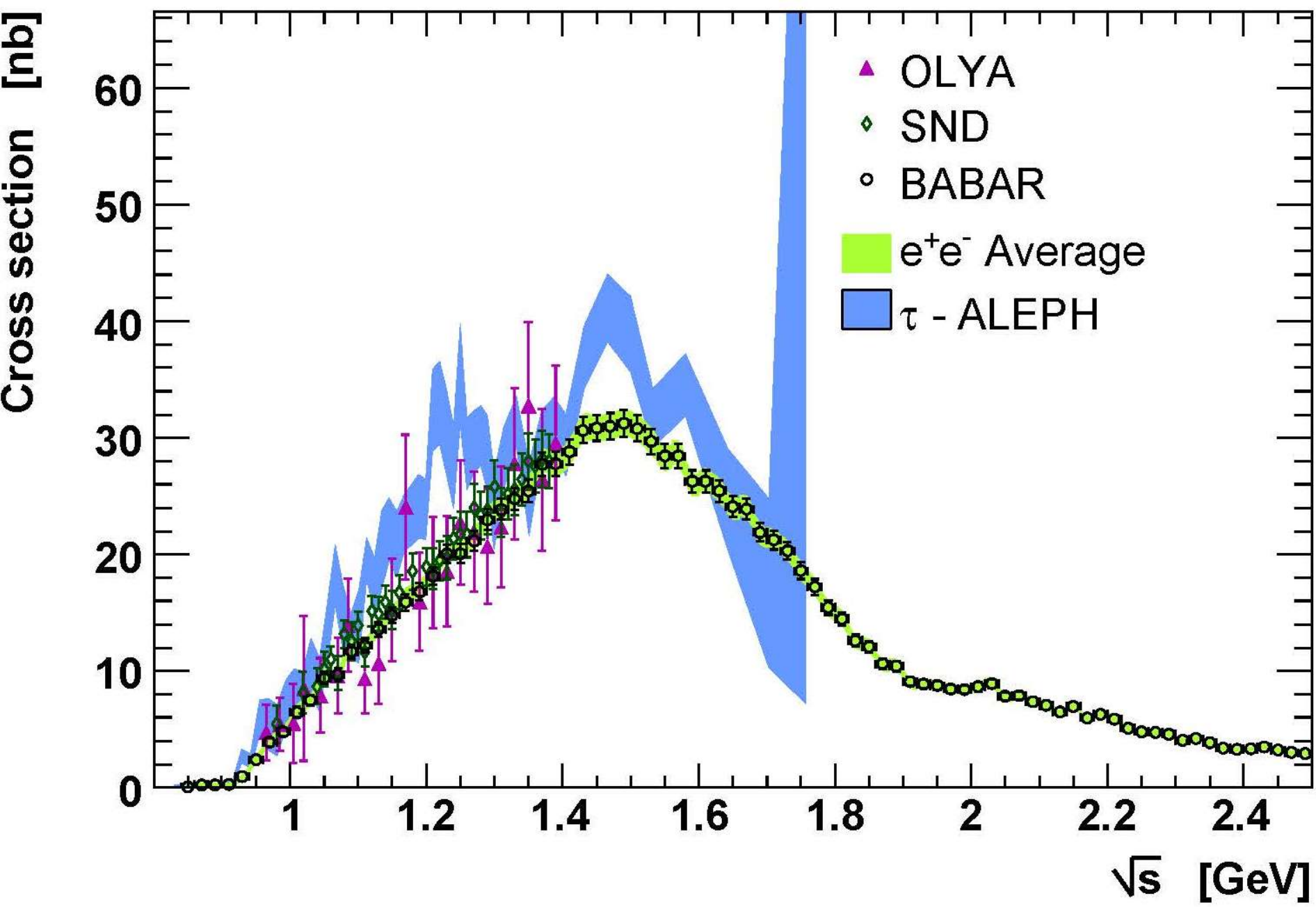}//
  \includegraphics[width=5cm]{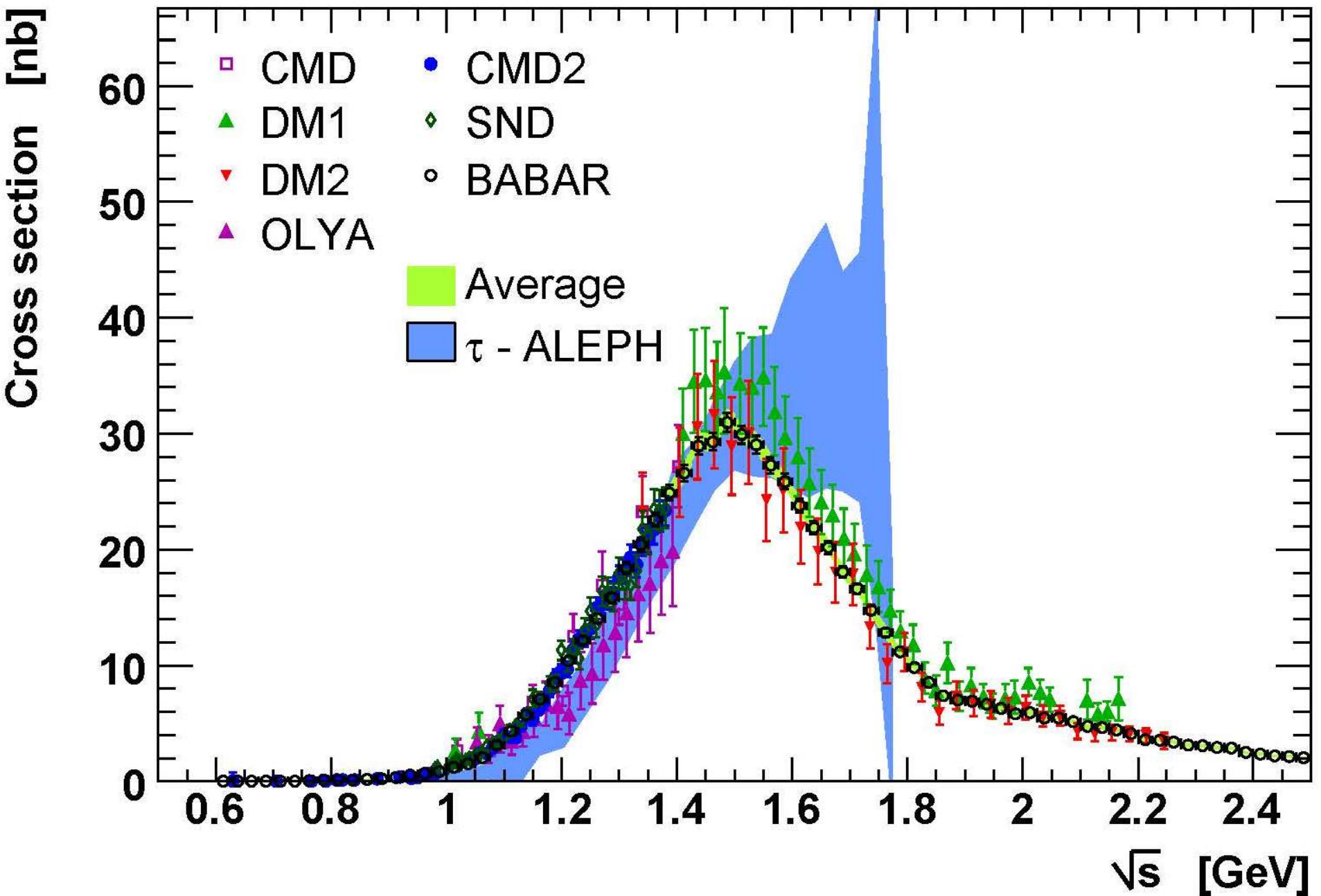}
  \vspace{-0.4cm}
  \caption{\small
The measured bare cross sections for $e^+e^-\to \pi^+ \pi^- 2\pi^0(\gamma)$ (top)
and $e^+e^-\to 2\pi^+ 2\pi^- (\gamma)$ (bottom)
with the combination band from threshold to 2.5 GeV. The corresponding predictions
from ALEPH $\tau$ spectral functions (darker band) are superimposed.}
  \label{tau-ee}
\end{figure}

For the $2\pi^+ 2\pi^-$ channel new data with the full \babar\ sample were 
published~\cite{4pi-babar} 
in 2012 with 5 times more statistics and a smaller systematic uncertainty (2.4\%).
The resulting combined HVP contribution is now
$13.70 \pm 0.03_{\rm stat} \pm 0.28_{\rm uncor syst} \pm 0.13_{\rm cor syst}$ 
with a reduced total uncertainty (0.31) compared to the 2011 value (0.53). 
The ALEPH $\tau$ estimate~\cite{new-aleph-tau}, 
$12.79 \pm 0.65_{\rm exp} \pm 0.35_{\rm IB}$, is consistent, but much less 
precise. Since the $\tau$ estimate for the two four-pion channels have some
anticorrelation from the $\nu_\tau \pi^{\pm} 3\pi^0$ mode through the
isospin relations, it makes sense to combine the two channels. The $\tau$ value,
$33.81 \pm 1.53$, is then consistent with the corresponding $e^+e^-$ value,
$31.86 \pm 0.64$, within 1.2~$\sigma$. While there has been a steady progress in 
$e^+e^-$ results over the last two decades, it is disappointing that similar
advances have not occured for the $\tau$ spectral functions. The situation
is illustrated in Fig.~\ref{tau-ee}.

\section{The channels $K \overline{K}$}
\label{KKbar}

\begin{figure}[htb]
  \centering
%  \vspace{-0.5cm}
  \includegraphics[width=7.5cm]{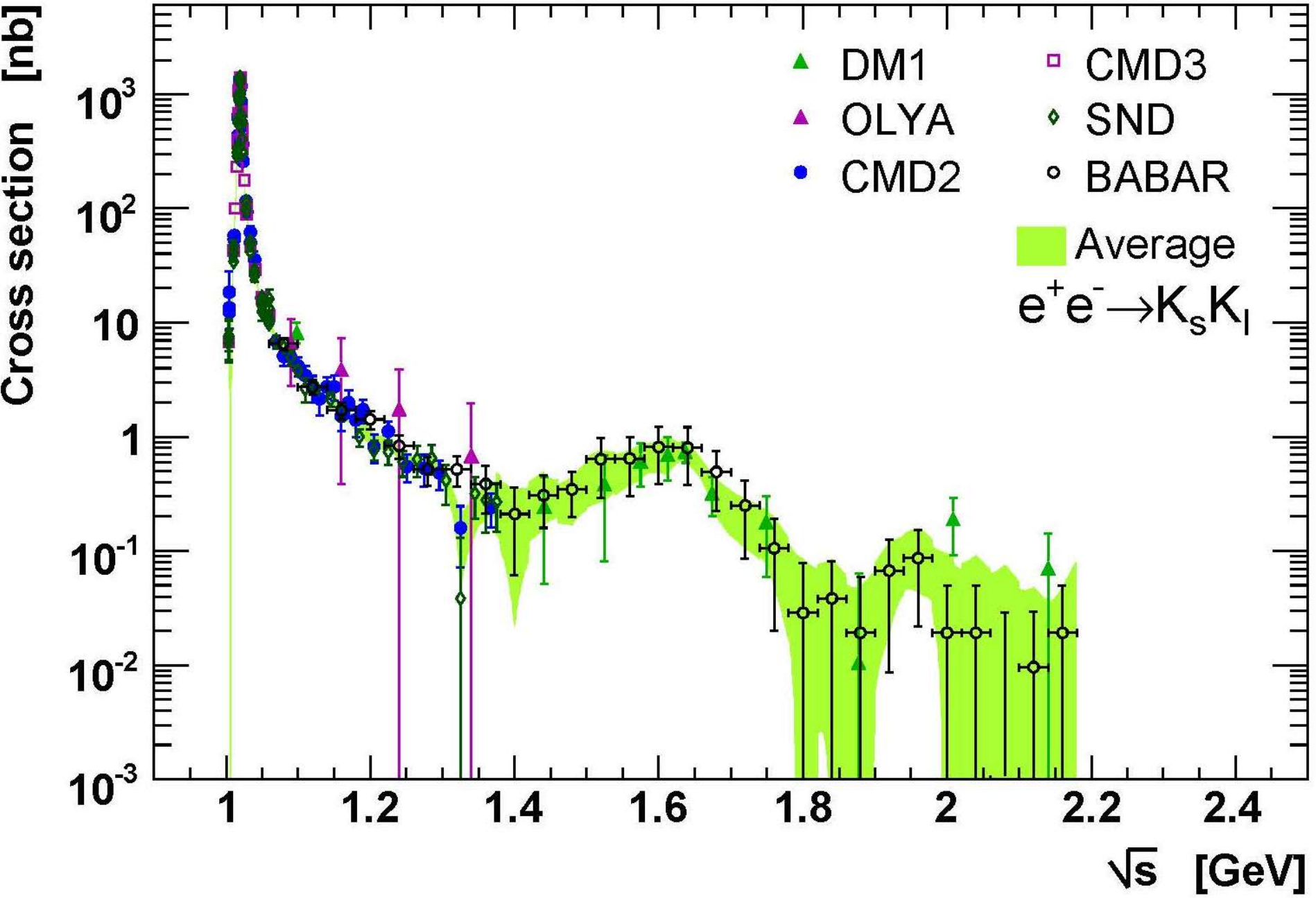}
  \vspace{-0.4cm}
  \caption{\small
The measured bare cross sections for $e^+e^-\to K_s K_l (\gamma)$  
with the combination band from threshold to 2.2 GeV.}
  \label{kskl}
\end{figure}

New data are available for the $K_sK_l$ channel: \babar~\cite{babar-kskl}
detects both $K_s$ and $K_l$ from threshold to 2.2 GeV, while 
CMD-3~\cite{cmd3-kskl} measures only $K_s$ in the $\phi$ resonance region.
A good consistency is observed for the $\phi$ between the two experiments as
well as with older ones (CMD-2 and SND). The cross section is given in
Fig.~\ref{kskl}.

The new $K_sK_l$ contribution to $a_\mu^{\rm had~LO}$ up to 1.8 GeV is
$12.81 \pm 0.06_{\rm stat} \pm 0.18_{\rm uncor syst} \pm 0.15_{\rm cor syst}$ with
a total uncertainty (0.24) reduced from the 2011 value (0.39).

\begin{figure}[htb]
  \centering
%  \vspace{-0.5cm}
  \includegraphics[width=7.5cm]{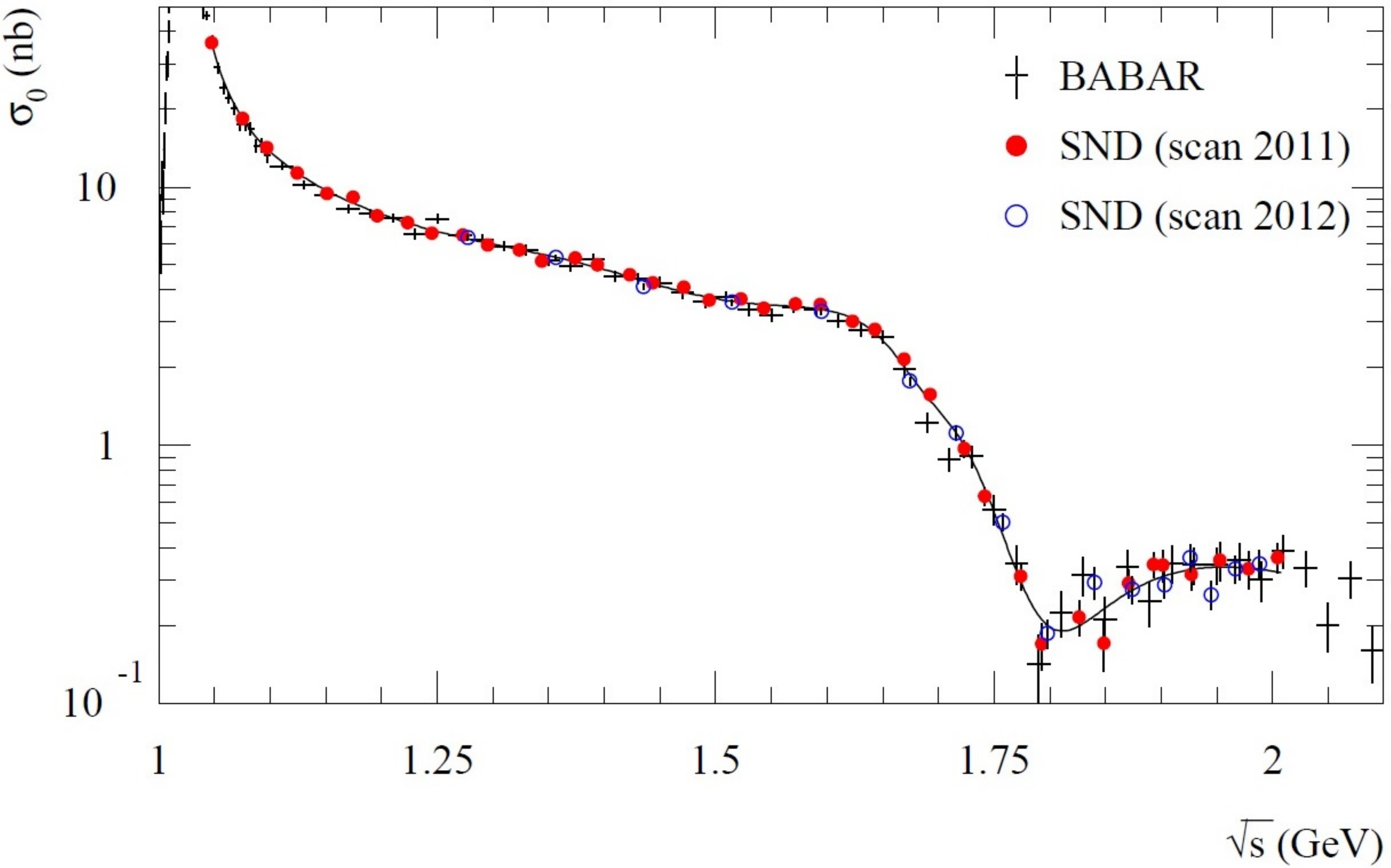}
  \vspace{-0.4cm}
  \caption{\small
The measured bare cross sections for $e^+e^-\to K^+ K^- (\gamma)$ from \babar\ 
and SND at VEPP-2000.}
  \label{kpkm}
\end{figure}

Recent results from SND~\cite{snd-kpkm} at VEPP-2000 for the $K^+K^-$ channel
agree well with \babar~\cite{babar-kpkm}, while both show a discrepancy with
the former SND results at VEPP-2M below 1.4 GeV beyond the quoted systematic
uncertainty. The \babar\ and the new SND data are displayed in Fig.~\ref{kpkm}.

Some concern is arising on the $e^+e^-\to\phi\to K^+K^-$ cross section. The \babar\
result has a systematic uncertainty of 0.7\%, but it is higher by 5.1\% (9.6\%) 
with respect to CMD-2 (SND) with respective systematic uncertainties of 2.2\% 
(7.1\%). Including the \babar\ data the new contribution to $a_\mu^{\rm had~LO}$
increases from 21.63 to 22.67 with a new uncertainty of 0.43. However a new
preliminary result from CMD-3~\cite{cmd3-kpkm} shows a huge increase ($\sim$11\%) with
respect to CMD-2, $\sim$5\% now above \babar! It should be noted in this respect that
the ISR method is more reliable than the scan technique for the detection of the 
slow charged kaons in the $\phi$ system because the kaons are boosted.

\section{The channels $K \overline{K} + pions$}
\label{KKbar-pions}

For our previous analyses the available data on $e^+e^-\to K\overline{K}+n{\rm pions}$
did not cover all the channels. Fortunately, it was possible to partially overcome this
lack of information by using constraints based on the knowledge of final-state dynamics 
and isospin symmetry~\cite{dehz2003-1,dhmz2010}. This procedure is now becoming unnecessary due 
to the release of new results from the systematic measurements of exclusive processes by 
\babar. 

\begin{figure}[htb]
  \centering
%  \vspace{-0.5cm}
  \includegraphics[width=5cm]{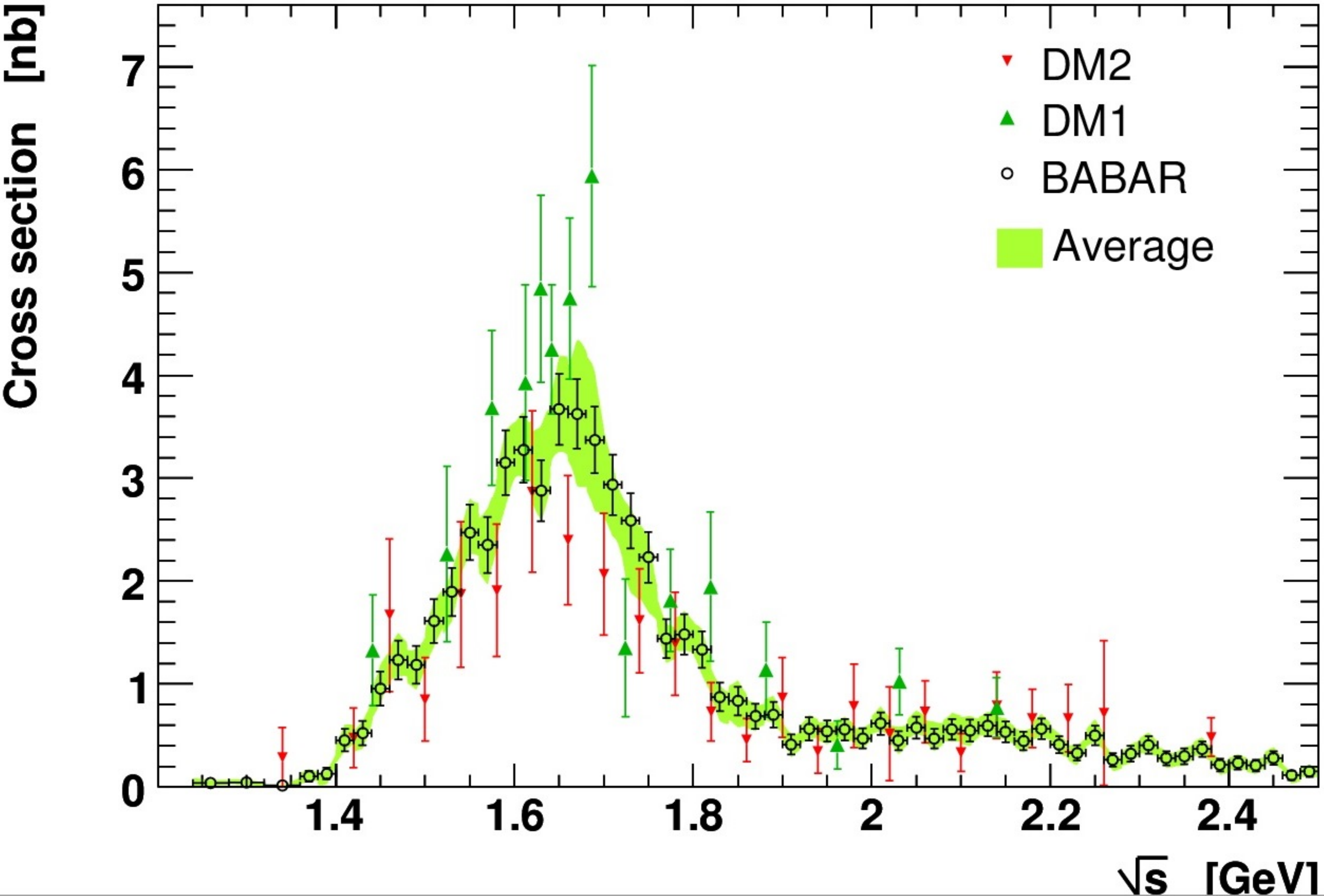}//
  \includegraphics[width=5cm]{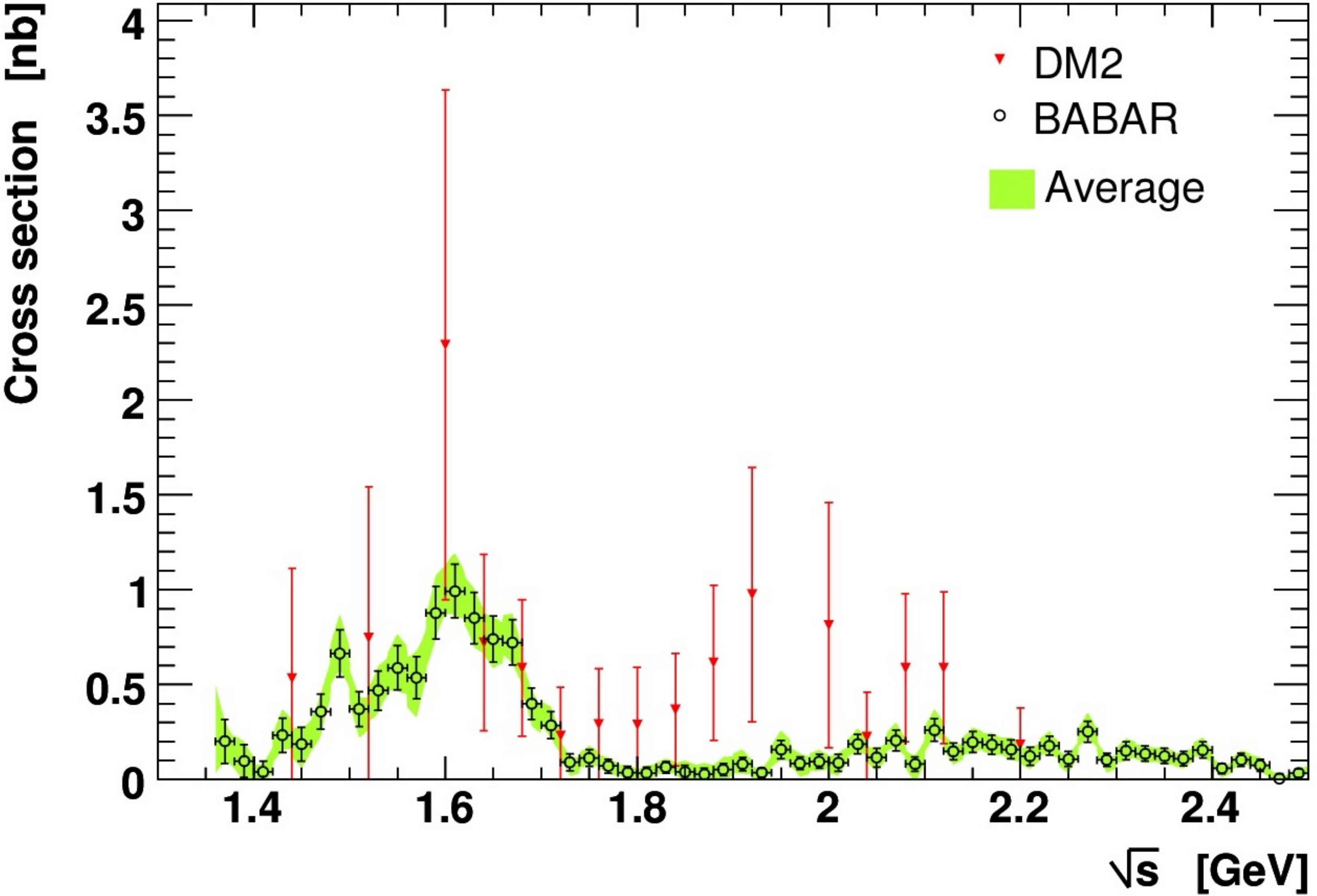}//
  \includegraphics[width=5cm]{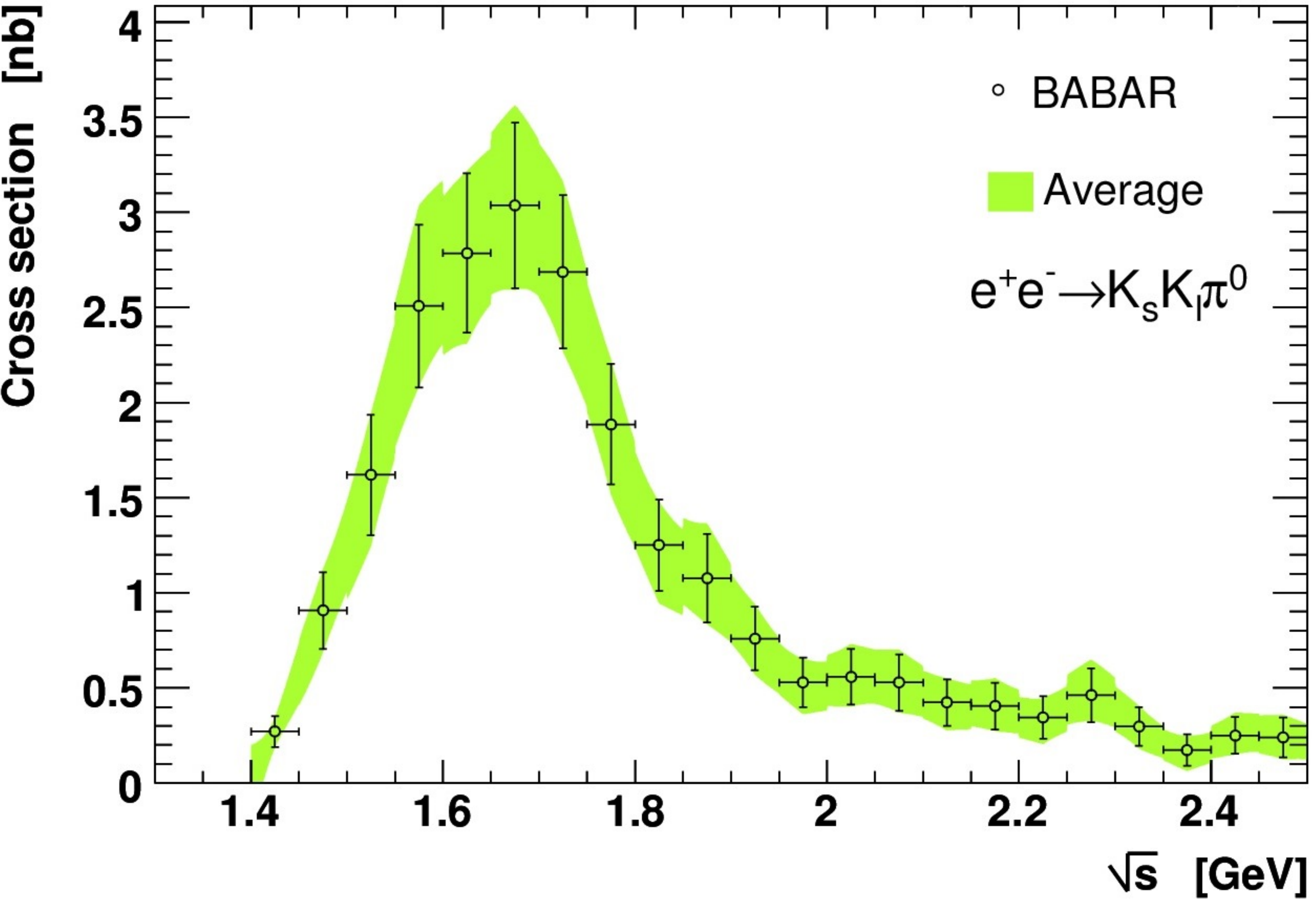}
  \vspace{-0.4cm}
  \caption{\small
The measured bare cross sections for all $e^+e^-\to K \overline{K} (\gamma)$ channels 
from \babar\ and older experiments.}
  \label{kkbarpi}
\end{figure}

Together with previous measurements of $K_s K^{\pm} \pi^{\mp}$ and $K^+K^- \pi^0$, data 
on the $K_sK_l \pi^0$ channel~\cite{babar-ksklpi0} now complete the picture for
$n=1$ (Fig.~\ref{kkbarpi}). The $K \overline{K} \pi$ final states being overwhelmingly 
dominated by $K^*(890) \overline{K} +cc$ below 1.8 GeV (with a small contribution 
from $\phi \pi^0$), it is not surprising that the isospin procedure works well. 
Indeed, the contribution from the sum of the measured channels is  $2.45 \pm 0.15$, 
in agreement (with similar precision) with the value $2.39 \pm 0.16$ using only 
$K_s K^{\pm} \pi^{\mp}$ data and isospin constraints.

For $n=2$ many channels contribute, of which only two ($K^+K^-\pi^+\pi^-$ and 
$K^+K^- 2\pi^0$) were known in 2010 from \babar. Thus isospin constraints were used, 
but since the dynamics is more complicated here with final states 
$K^*(890) \overline{K} \pi +cc$, $K \overline{K} \rho$ and $\phi \pi \pi$, the
systematic uncertainty had to be enlarged. New modes have now been added 
and older ones updated: 
$K^+K^-\pi^+\pi^-$ and $K^+K^- 2\pi^0$~\cite{babar-kkpipi}, $K_sK_l\pi^+\pi^-$ and 
$K_sK_s \pi^+\pi^-$~\cite{babar-kskl}, and $K_sK_l 2\pi^0$~\cite{babar-ksklpi0}. 
Apart for the $K_lK_l \pi^+\pi^-$ channel which 
can be safely estimated using CP symmetry, all the cross sections have been measured.
The only channel not yet released is $K_s K^{\pm} \pi^{\pm} \pi^0$, to appear shortly.
The expected precision on the contribution from all $n=2$ modes is 0.06, a large step
from the previous isospin systematic uncertainty of 0.39 which is still kept until 
the final \babar\ release.

\section{Updated Standard Model prediction}
\label{update}

Taking into account all the updated contributions our preliminary 2016 value for
$a_\mu^{\rm had~LO}$ becomes

\begin{equation}
\label{res-2016}
 a_\mu^{\rm had~LO}=692.6 \pm 1.2 \pm 2.6 \pm 1.6 \pm 0.1 \pm 0.3~~~~~~~(2016)~.
\end{equation}
where the uncertainties are statistical, systematic from uncommon and common sources,
from $\psi$, and QCD, respectively.

The central values for 2016 and 2011 Eq.~(\ref{res-2010}) are in agreement. However
the total uncertainty is significantly reduced from 4.2 to 3.3 (21\%). Also the result
shows the importance to take into account properly the systematic uncertainties which 
are correlated between channels for a given experiment and also between experiments.

Putting now together all the contributions to $a_\mu$, 
$a_\mu^{\rm QED}=11658471.895 \pm 0.008$~\cite{amu-qed},
$a_\mu^{\rm EW}=15.4 \pm 0.1$~\cite{amu-ew},
$a_\mu^{\rm had~LBL}=10.5 \pm 2.6$~\cite{amu-lbl},
$a_\mu^{\rm had~LO}=692.6 \pm 3.3$,
$a_\mu^{\rm had~NLO}=-9.87 \pm 0.09$~\cite{amu-hadnlo},
$a_\mu^{\rm had~NNLO}=1.24 \pm 0.01$~\cite{amu-hadnlo},
we obtain $a_\mu^{\rm SM~prediction}=11659181.7 \pm 4.2$ to be compared to the
direct measurement~\cite{bnl} $a_\mu^{\rm exp}=11659209.1 \pm 6.3$. Their
difference, $27.4 \pm 7.6$, remains at the 3.6~$\sigma$ level, the reduction of the
prediction uncertainty being compensated by the inclusion of the recently 
calculated NNLO hadronic contribution.

\begin{figure*}[htb]
  \centering
%  \vspace{-0.5cm}
  \includegraphics[width=12cm]{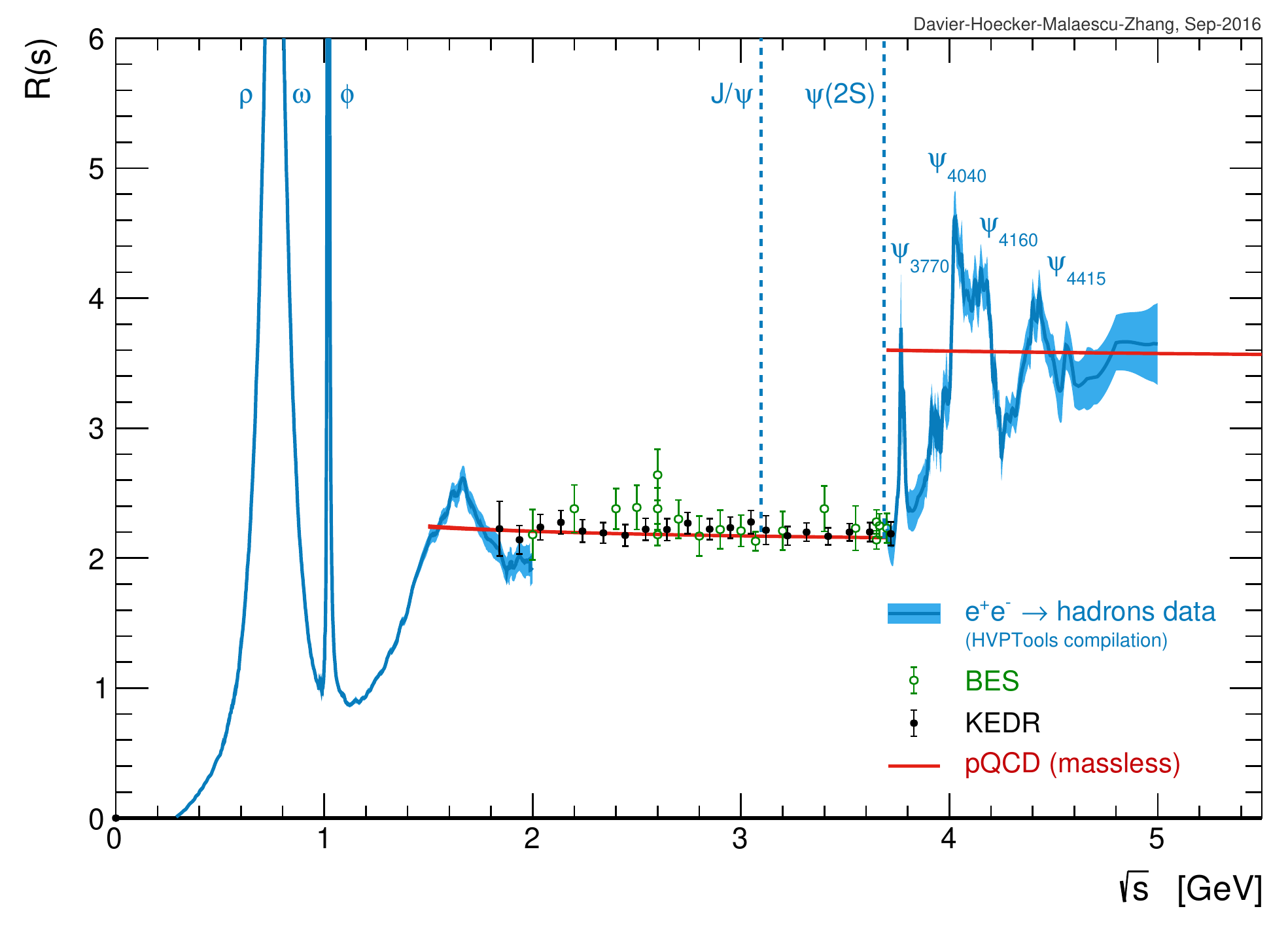}
  \vspace{-0.4cm}
  \caption{\small
The tatal hadronic annihilation rate $R$ as a function of $\sqrt{s}$. Inclusive 
measurements from BES~\cite{bes-r} (and references therein) 
and KEDR~\cite{kedr-r-1,kedr-r-2} are given as data points,
while the sum of exclusive channels from this analysis is given by the narrow 
bands.}
%below 2 GeV and from 3.77 to 5 GeV.}
  \label{R}
\end{figure*}

The nearly complete set of exclusive cross sections from \babar, complemented by 
results from other experiments for some channels allows one to compute the total
$e^+e^-$ annihilation rate to hadrons $R(s)$, expressed in units of the point-like 
pair cross section. This is the most accurate determination to date which can be 
trusted up to 2 GeV. Clearly at larger $\sqrt{s}$ values many more exclusive 
channels open up and inclusive $R$ measurements are necessary. In this respect, 
as shown in Fig.~\ref{R} the newly published results from KEDR~\cite{kedr-r-2} 
between 1.84 and 3.05 GeV, complementing previous ones between 3.12 and 
3.72 GeV~\cite{kedr-r-1} and the BES results~\cite{bes-r}, 
overlap nicely with our compilation and show excellent 
agreement with perturbative QCD, limiting serious deviations from local 
quark-hadron duality. The overall strongly damped oscillatory behaviour of $R(s)$ 
around the QCD prediction justifies using energy-averaged (global) quark-hadron 
duality~\cite{svz} as a reliable tool to estimate dispersion integrals 
in the nearly-asymptotic regime.

\section{Conclusion and perpectives}
\label{conclu}

Using all available data on the $e^+e^-\to {\rm hadrons}$ cross sections an
update of the lowest-order hadronic vacuum polarisation to the muon magnetic
anomaly is obtained with a relative precision of 0.5\%: 
$a_\mu^{\rm had~LO}=(692.6 \pm 3.3)\times 10^{-10}$.
The achieved uncertainty on this contribution is now reduced to about 
half the current uncertainty of the direct $a_\mu$ measurement.
Thus the forthcoming programs at Fermilab~\cite{fnal} and JPARC~\cite{jparc}, 
aiming at a precision four times smaller are therefore necessary to confirm 
if the present 3.6~$\sigma$ deviation is due to new physics beyond the 
Standard Model.

In order to match the precision of the future direct measurements, 
experimental progress is still needed to reduce further the uncertainty on 
$a_\mu^{\rm had~LO}$ from dispersion relations. Analyses for the $\pi^+\pi^-$ 
channel are underway with \babar\ using a new independent method and CMD-3, 
where a systematic uncertainty of 0.3\% looks reachable. In the 1-2 GeV range it 
will be important to continue to confront the \babar\ and CMD-3/SND results.
Independently, lattice
calculations are also progressing, but they are not yet at a competitive level.

The precision of $a_\mu^{\rm had~LO}$ (3.3) is now getting close to the estimated
systematic uncertainty on the hadronic LBL contribution $a_\mu^{\rm had~LBL}$ (2.6) 
which appears for the moment irreducible. Here only models have been used so far and 
lattice calculations are badly needed. However it should be pointed out that even 
if the LBL systematic uncertainty stays at the present level, the combined progress
of $e^+e^-$ data for the prediction and of the direct measurements would be sufficient 
to boost the present deviation (if persisting) to a level of 7~$\sigma$, thereby allowing
one to unambiguously claim a breakdown of the Standard Model.

I would like to thank A.~Hoecker, B.~Malaescu and Z. Zhang for the fruitful
collaboration and our IHEP friends for organising a perfect $\tau$ 
workshop.



%% References

%*\nocite{*}
%*\bibliographystyle{elsarticle-num}
%*\bibliography{martin}



\end{document}